\documentstyle[12pt]{article}

\begin{document}

\begin{flushright}
IMSc/2008/07/12 
\end{flushright} 

\vspace{2mm}

\vspace{2ex}

\begin{center}

{\large \bf Stabilisation of Seven Directions in an \\



\vspace{2ex}

Early Universe M -- theory Model
}

\vspace{8ex}

{\large  Samrat Bhowmick, Sanatan Digal, 
\footnote{Corresponding author}
and S. Kalyana Rama}

\vspace{3ex}

Institute of Mathematical Sciences, C. I. T. Campus, 

Tharamani, CHENNAI 600 113, India. 

\vspace{1ex}

email: samrat, digal, krama@imsc.res.in \\ 

\end{center}

\vspace{6ex}

\centerline{ABSTRACT}
\begin{quote} 

Our model consists of intersecting $22'55' \;$ branes in M
theory distributed uniformly in the common transverse space.
Equations of state follow from U duality symmetries. In this
model, three spatial directions expand, and seven directions
stabilise to constant sizes. From string theory perspective, the
dilaton is hence stabilised. The constant sizes depend on
certain imbalance among initial values. One naturally obtains
$M_{11} \simeq M_s \simeq M_4 \;$ and $g_s \simeq 1 \;$ within a
few orders of magnitude. Smaller numbers, for example $M_s
\simeq 10^{- 16} \; M_4 \; $, are also possible but require fine
tuning.

\end{quote}

\vspace{2ex}


\newpage


\vspace{4ex}

{\bf 1.}
In early universe, the temperature and energy densities are
high. When they are of the order of Planck scale $M_4 \simeq
10^{19} GeV \;$, the dynamics of the early universe is expected
to be described by a more fundamental theory such as string
theory or M theory \cite{bowick, k2}.

If this is the case then the problem of spacetime dimensions
needs to be resolved -- spacetime is eleven dimensional in M
theory whereas it is four dimensional in our observed universe.

A canonical resolution is that the early universe starts out
being eleven dimensional. During its evolution, by some
dynamics, seven of the spatial directions cease to expand and
their sizes become stabilised. The remaining three spatial
directions continue to expand and become the observed universe.

The stabilised sizes then relate the M theory scale $M_{11} \;$
and the four dimensional Planck scale $M_4 \;$. Likewise, since
string theory can be obtained by dimensionally reducing M
theory, the sizes also relate $M_{11} \;$ and the string scale
$M_s \;$ and the string coupling constant $g_s \;$. One may then
enquire, for example, whether it is possible to have string/M
theory scale in the $TeV$ range as required in Large Volume
compactification scenarios \cite{add}.

Various proposals have been made for obtaining a four
dimensional universe from string/M theory \cite{bv, gas, 3+1}.
Typically, one assumes that the spatial directions are all
toroidal, and are wrapped by a gas of winding and anti winding
strings or $p$--branes; and that the cosmological evolution is
governed by a ten/eleven dimensional effective action. The
earliest proposal \cite{bv}, in the context of string theory, is
based on the observation that winding and anti winding strings
oppose the expansion, and are annihilated efficiently in four
dimensional spacetime. Others \cite{gas, 3+1} are variants of
this, or based on its generalisations to winding and anti
winding $p$--branes in string/M theory. These proposals are
quite appealing and have been used in a variety of `brane gas'
models \cite{gas, 3+1}, but some important issues yet remain to
be resolved \cite{gm}.

In this letter, based on the ideas in \cite{cm, k}, we present
an M theoretic early universe model where seven of the spatial
directions cease to expand and their sizes become stabilised.
From string theory perspective, the dilaton is hence stabilised.
The remaining three spatial directions continue to expand, thus
leading to a four dimensional universe. The stabilised sizes,
and thus the explicit relations among $(M_{11}, \; M_4, \; M_s,
\; g_s) \;$, depend on certain imbalance among initial values.
The exact values are obtained numerically, but can also be
estimated analytically under certain approximations. In this
model, one may obtain any value for $M_{11} \;$ or $M_s \;$,
including in the $TeV$ range, by a corresponding fine tuning of
initial values.

\vspace{4ex}

{\bf 2.}
Our model is as follows. Let all the spatial directions be
toroidal. Consider mutually BPS intersecting brane
configurations in M theory where $N$ sets of coincident branes
and antibranes intersect as per the rules given in \cite{bps}.
According to these rules, for example, two sets of 2 branes must
intersect along zero common direction, 2 branes and 5 branes
along one common direction, or two sets of 5 branes along three
common directions.

The branes and antibranes in such a configuration differ
significantly from those in brane gas models, as explained in
section 2.6 of the first and section 6 of the second paper in
\cite{cm}. Briefly, the differences are the following: (1) In
brane gas models, the branes can intersect each other
arbitrarily. Here the intersections must follow specific
rules. U duality symmetries of M theory then imply a relation
among the equations of state which turns out to be a crucial
element underlying the present results \cite{k}. (2) The branes
in brane gas models support excitations on their surfaces and,
at high energies, have $S \sim {\cal E} \;$ where $S$ is the
entropy and ${\cal E}$ the energy. Here, the intersecting branes
form bound states, become fractional, support very low energy
excitations and, hence, are highly entropic. At high energies,
$S \sim {\cal E}^{\frac{N} {2}} \;$ which, for $N > 2 \;$,
vastly exceeds the entropy in brane gas models. Such
intersecting brane configurations are, therefore, the
entropically favourable ones. (3) In brane gas models, the
branes tend to annihilate if they intersect each other. Here,
the intersections are necessary for formation of bound states
and of high entropic excitations.  These excitations are
long--lived and non interacting to the leading order, hence the
branes here are metastable and do not immediately annihilate.
See \cite{cm} for more details, and \cite{hm} also.

In our model, we consider $N = 4 \;$ intersecting brane
configuration denoted by $22'55' \;$, which has vanishing net
charges and consists of two sets each of $2 \;$ branes and $5
\;$ branes along $(x^1, x^2) \;$, $(x^3, x^4) \;$, $(x^1, x^3,
x^5, x^6, x^7) \;$, and $(x^2, x^4, x^5, x^6, x^7) \;$
directions. This configuration, when localised in the common
transverse space along $(x^8, x^9, x^{10}) \;$ directions,
describes a four charged black hole \cite{cvetic}. Here, we take
the configuration to be uniformly distributed in the common
transverse space which then is assumed, as in \cite{cm, k}, to
describe a homogeneous anisotropic universe whose evolution is
governed by an eleven dimensional effective action.

Let $I = 1, 2, 3, 4 \;$ denote the branes $2, 2', 5, 5' \;$
respectively. We assume that, as in the case of black holes, the
energy momentum tensors $T^A_{\; \; B (I)} \;$ of the $I^{th}$
set of branes are mutually non interacting and seperately
conserved \cite{cm, hm}. Then
\begin{equation}\label{tab}
T^A_{\; \; B} = \sum_I T^A_{\; \; B (I)} \; \; , \; \; \;
\sum_A \nabla_A \; T^A_{\; \; B (I)} = 0 
\end{equation}
where $T^A_{\; \; B}$ is the total energy momentum tensor of the
configuration. Homogeneity implies that $T^A_{\; \; B} = diag \;
(- \rho, p_i) $ and $T^A_{\; \; B (I) } = diag \; (- \rho_I,
p_{i I}) \;$. We take $\rho_I > 0 \;$.

To obtain the equations of state $p_{i I} (\rho_I) \;$, let
$p_{\parallel I} \;$ and $p_{\perp I} \;$ denote parallel and
perpendicular components of pressure due to $I^{th} \;$ set of
branes. For the mutually BPS intersecting brane configurations
of the type considered here, it is shown in \cite{k} that U
duality symmetries of M theory imply that the functions
$p_{\perp I} (\rho_I) \;$ must be same for all $I$ and that
$p_{\parallel I} = 2 p_{\perp I} - \rho_I \;$. For the $22'55'
\;$ configuration, it then folows that if $\rho_I \;$ are all
equal then, for any function $p_\perp (\rho) \;$, the seven
brane directions become stabilised and the remaining three
spatial directions expand \cite{k}.

However, an explicit form for the function $p_\perp (\rho) \;$
is required to obtain further details such as the values of the
stabilised sizes, or to understand the evolution when $\rho_I
\;$ are all not equal. In principle, $p_\perp (\rho) \;$ is to
be determined by brane antibrane dynamics. But not much is known
about this dynamics. Hence, in order to make progress and to
understand the details of the evolution, we assume in our model
that $p_\perp = (1 - u) \; \rho \;$ where $u \;$ is a constant.
Such a form, with $u = 1 \;$, is indeed derived in \cite{cm} in
the limit where the brane antibrane annihilation can be
neglected. Here, we will keep $u$ an arbitrary constant,
assuming only that $0 < u < 2 \;$. The resulting evolution is
then applicable, atleast qualitatively, even if $u$ is varying
{\em e.g.} due to brane antibrane annihilation effects.

It then follows that $p_{i I} = (1 - u^I_i) \; \rho_I \;$ where,
for the $22'55' \;$ configuration,
\begin{eqnarray} 
u_i^1 & = & u \; (2, 2, 1, 1, 1, 1, 1, 1, 1, 1) \nonumber \\
u_i^2 & = & u \; (1, 1, 2, 2, 1, 1, 1, 1, 1, 1) \nonumber \\
u_i^3 & = & u \; (2, 1, 2, 1, 2, 2, 2, 1, 1, 1) \nonumber \\
u_i^4 & = & u \; (1, 2, 1, 2, 2, 2, 2, 1, 1, 1) \; \; . 
\label{I} 
\end{eqnarray}

\vspace{4ex}

{\bf 3.}
Consider now the evolution of the $D = (10 + 1)$ -- dimensional
homogeneous anisotropic universe in the model described above.
Let the line element $d s \;$, with $x^A = (t, \; x^i) \;$ and 
$i = 1, 2, \cdots, D - 1 \;$, be given by
\begin{equation}\label{ds}
d s^2 = \sum_{A B} g_{A B} d x^A d x^B = - d t^2 
+ \sum_i e^{2 \lambda^i} \; (d x^i)^2 
\end{equation}
where $\lambda^i$ are functions of $t$ only. Einstein equations
$R_{A B} - \frac{1}{2} g_{A B} R = T_{A B} \;$, with $8 \pi G =
1$, and equations (\ref{tab}) lead to $\rho_I = e^{l^I - 2
\Lambda} \;$ and
\begin{eqnarray} 
\sum_{i j} G_{i j} \lambda^i_t \lambda^j_t & = & 
2 \; \sum_I e^{l^I - 2 \Lambda} \label{quad2} \label{11} \\
\lambda^i_{t t} + \Lambda_t \lambda^i_t & = & 
\sum_I \; u^{i I} \; e^{l^I - 2 \Lambda}  
\label{12} 
\end{eqnarray}
where $l^I = \sum_i u^I_i \lambda^i + l^I_0 \;$, $\; \Lambda =
\sum_i \lambda^i \;$, the subscripts $t$ denote time
derivatives, and
\begin{equation}
G_{i j} = 1 - \delta_{i j} \; \; , \; \; \; 
G^{i j} = \frac{1}{D - 2} - \delta^{i j} \; \; , \; \; \;
u^{i I} = \sum_j G^{i j} u^I_j  \; \; . \label{Gij} 
\end{equation}
Let $d \tau = e^{- \Lambda} \; dt \;$ and ${\cal G}^{I J} =
\sum_i u^{i I} u^J_i \;$. Also, define ${\cal G}_{I J} \;$ by
$\sum_J {\cal G}^{I J} \; {\cal G}_{J K} = \delta^I_K \;$.
Then, manipulating equations (\ref{11}) and (\ref{12}), one
obtains
\begin{eqnarray} 
\lambda^i & = & \sum_{I J} {\cal G}_{I J} \; u^{i I} \; 
( l^J - l^J_0 ) + L^i \; \tau \label{lia} \\
l^I_{\tau \tau} & = & \sum_J {\cal G}^{I J} \; e^{l^J} 
\label{42a} \\
\sum_{I J} {\cal G}_{I J} \; l^I_\tau l^J_\tau & = & 
2 \; ( E + \sum_I e^{l^I} ) \label{41a} 
\end{eqnarray}
where the subscripts $\tau$ denote $\tau$--derivatives, $L^i \;$
are integration constants satisfying $\sum_i u^I_i L^i = 0 \;$,
and $2 \; E = - \sum_{i j} G_{i j} L^i L^j \;$. Also, with no
loss of generality, we have taken the initial values to be
\begin{equation}\label{ic}
\left( \lambda^i , \; \lambda^i_t , \; l^I, \; l^I_t,
\; \rho_I , \; \tau \right)_{t = 0} =
\left( 0, \; k^i, \; l^I_0, \; K^I, \; \rho_{I 0}, \; 0 \right)
\end{equation}
where $\rho_{I 0} = e^{l^I_0} \; $ and $k^i = \sum_{I J} {\cal
G}_{I J} \; u^{i I} \; K^J + L^i \; $. For the $22'55' \;$
configuration in our model, $u^I_i \;$ are given in equations
(\ref{I}) using which $u^{i I} \;$, $\; \; {\cal G}^{I J} \;$,
and $ {\cal G}_{I J} \;$ can be calculated easily. For example,
\begin{equation}\label{neata}
{\cal G}^{I J} = 2 u^2 \; \left( 1 - \delta^{I J} \right)
\; \; \; , \; \; \;  \; \; \; 
{\cal G}_{I J} = \frac{1}{6 u^2} \; ( 1 - 3 \delta_{I J} ) 
\; \; . 
\end{equation}

We now point out an interesting similarity with black holes:
When $L^i \;$ all vanish, $e^{\lambda^i} \;$ here have the same
form as those for extremal $22'55' \;$ black holes and $e^{2 u
h_I} \;$, where $h_I = \sum_J {\cal G}_{I J} \; ( l^J - l^J_0 )
\;$, play the role of harmonic functions $H_I = 1 + \frac{Q_I}
{r} \;$. Compare equations (\ref{lia}) here and (18) in
\cite{cvetic}. Also, the asymptotic limit $t \to \infty \;$
here, see below, corresponds to the near horizon limit $r \to 0
\;$ and (certain combination of) $\rho_{I 0} \;$ play the role
of $Q_I \;$.


\vspace{4ex}

{\bf 4.}
To obtain $\lambda^i (t) \;$ for the $22'55' \;$ configuration,
and thus the evolution of the universe, one may solve equations
(\ref{42a}) -- (\ref{neata}) for $l^I (\tau) \;$ and obtain
$\lambda^i (\tau) \;$ from equation (\ref{lia}). Then $t (\tau)
\;$ and, hence, $\tau (t) \;$ follow from $d t = e^\Lambda \; d
\tau \;$. We are unable to solve equations (\ref{42a}) --
(\ref{neata}) analytically. Nevertheless, the important features
of the evolution can be obtained as follows.

For the $22'55' \;$ configuration, the following two results can
be proved: {\bf (R1)} The constraints $\sum_i u^I_i L^i = 0 \;$
imply that $0 \le c_i \; (L^i)^2 \le E \;$ where $c_i \;$ are
constants of ${\cal O} (1) \;$. Hence $E = 0 \;$ if and only if
$L^i = 0 \;$ for all $i \;$. {\bf (R2)} If $E \ge 0 \;$ then
equations (\ref{lia}) and (\ref{41a}) imply that none of
$(\Lambda_\tau , \; l^I_\tau) \;$ may vanish, and that they must
be {\em all positive} or {\em all negative}.

Let $K^I = l^I_\tau (0) > 0 \;$ for all $I \;$. The above
results together with equations (\ref{42a}) and (\ref{neata})
then imply that, as $\tau \;$ increases, $l^I (\tau) \;$ all
increase and diverge at finite $\tau = \tau_\infty \;$. In the
limit $\tau \to \tau_\infty \;$ and to the leading order, we
obtain
\begin{eqnarray}
e^{l^I} \; = \; \frac{1}{3 u^2} \; 
\frac{1}{(\tau_\infty - \tau)^2} \; \; \; , \; \; \; \;
t & = & t_* + A \; \left( \tau_\infty - \tau 
\right)^{- \; \frac{2 - u}{u}} \nonumber \\ 
e^{\lambda^i} \; = \; e^{v^i} \; \left( \frac{1}{3 u^2} \; \;
\frac{1}{(\tau_\infty - \tau)^2} \right)^{\sum_{I J} 
{\cal G}_{I J} \; u^{i J}} & = & e^{v^i} \; \; 
\left\{ B \; ( t - t_* ) \; \right\}^{\beta^i} \label{final}
\end{eqnarray}
where $\; t_* \;$ and $\tau_\infty \;$ are finite constants and
depend on the details of evolution, $A \;$ and $B \;$ are
$u$--dependent constants, $v^i = - \sum_{I J} {\cal G}_{I J} \;
u^{i I} \; l^J_0 + L^i \; \tau_\infty \;$, and $\beta^i =
\frac{2 u}{2 - u} \; \sum_{I J} {\cal G}_{I J} \; u^{i J}
\;$. Explicitly, $\beta^i \;$ are given by 
\begin{equation}
\beta^i \; = \; \frac{2}{3 (2 - u)} \; 
(0, \; 0, \; 0, \; 0, \; 0, \; 0, \; 0, \; 1, \; 1, \; 1) 
\; \; .  \label{betai}
\end{equation} 

Thus, asymptotically, $t \to \infty \;$ since $0 < u < 2 \;$ in
our model. And, $e^{\lambda^i} \to t^{\frac{2}{3 (2 - u)}} \; $
for the common transverse directions $i = 8, 9, 10 \;$. Hence,
these directions continue to expand, their expansion being
precisely that of a $(3 + 1)$ -- dimensional homogeneous,
isotropic universe containing a perfect fluid whose equation of
state is $p = (1 - u) \; \rho \;$. Also, $\; e^{\lambda^i} \to
e^{v^i} \;$ for the brane directions $i = 1, \cdots, 7
\;$. Hence, these directions cease to expand and their final
sizes are given by $e^{v^i} \;$.

In our model, irrespective of initial values, three spatial
directions will always expand and seven brane directions will
always be stabilised and reach constant sizes. The underlying
dynamics is distinct from those in \cite{bv, gas, 3+1} and can
be described as follows. It follows from equation (\ref{12})
that parallel brane directions contract and transverse ones
expand, at opposite rates for 2 branes and 5 branes. If the
brane energy densities $\rho_I \;$ are all different then,
generically, so will be the corresponding expansion and
contraction rates, and the brane directions will have net
expansion or contraction.  Only if the expansion rates equal
contraction rates, will the brane directions cease to expand or
contract and their sizes stabilise to constant values.

Such an equality ensues eventually in our model as a result of
two crucial features : {\bf (i)} The dynamics of the evolution,
given by $u^I_i \;$ which in turn follow from U duality
symmetries \cite{k}, is such that $\rho_I \;$, even if different
initially, evolve to become all equal. This equality is due to
each $\rho_I \sim e^{l^I} \;$ being `sourced' by the sum of
other three, see equations (\ref{42a}) and (\ref{neata}). {\bf
(ii)} The $22'55' \;$ configuration is such that each brane
direction is parallel to two sets of branes, and transverse to
other two in just the right way. Hence, its expansion and
contraction rates become equal once $\rho_I \;$ become all
equal.

The stabilised sizes of the brane directions should then depend
on the imbalance among $\rho_{I 0} \;$ and $\lambda^i_t (0)
\;$. Indeed we have, for example,
\begin{equation}\label{v1c}
e^{v^1} = e^{ L^1 \; \tau_\infty } \; \; \left( 
\frac{\rho_{2 0} \; \rho_{4 0}^2} {\rho_{3 0} \; \rho_{1 0}^2}
\right)^{\frac{1}{6 u}}
\; \; \; , \; \; \; \; 
e^{v^c} = e^{ L^c \; \tau_\infty } \; \; \left( 
\frac{\rho_{1 0} \; \rho_{2 0}} {\rho_{3 0} \; \rho_{4 0}}
\right)^{\frac{1}{6 u}} \; \; , 
\end{equation}
where we also define $v^c = \sum_{i = 1}^7 v^i \;$ and $L^c =
\sum_{i = 1}^7 L^i \;$, needed below. 

Thus, asymptotically as $t \to \infty \;$, the $(10 + 1)$ --
dimensional universe effectively becomes $(3 + 1)$ --
dimensional. Also, dimensional reduction of M theory along, for
example, $x^1 \;$ direction gives string theory with its dilaton
now stabilised. Let the coordinate sizes $\simeq {\cal O}(
\frac{1} {M_{11}}) \;$. Then, upto numerical factors of ${\cal
O} (1) \;$, the corresponding scales $(M_{11}, \; M_4, \; M_s)
\;$ and the string coupling constant $g_s \;$ are related
asymptotically by
\begin{equation}\label{411} 
M^2_4 \simeq \; e^{v^c} \; M^2_{11} 
\simeq e^{v^c - v^1} \; M_s^2 
\; \; \; , \; \; \;  \; \; \;  
g_s^2 \simeq e^{3 v^1 } \; \; . 
\end{equation} 

\vspace{4ex}

{\bf 5.}
To determine the sizes of brane directions and the relations in
equation (\ref{411}) explicitly for a given set of initial
values $(l^I_0, \; K^I, \; L^i) \;$, we need $\tau_\infty \;$ if
$L^i \ne 0 \;$. We will obtain $\tau_\infty \;$ numerically
since it depends on the details of evolution and we do not have
explicit solutions. But we first give an approximate expression
for $\tau_\infty \;$ which is easy to evaluate and works well
under certain conditions.

Let $L^i \ne 0 \;$. We set $E = 1 \;$ by measuring $t \;$ and
$\tau \;$ in units of $\frac{1}{\sqrt{E}} \;$. Note that if
$e^{l^I_0} \ll 1 \;$ for all $I \;$ then equations (\ref{42a})
and (\ref{41a}) imply that $l^I (\tau)$ may be taken as evolving
`freely', {\em i.e.}  $\; l^I (\tau) = l^I_0 + K^I \tau \;$
where $K^I = l^I_\tau (0) \; > 0 \;$, until one of the $e^{l^I}
= 1 \;$; from then on, all $e^{l^I}$ will evolve quickly and
diverge soon after. Consequently, $\tau_\infty \;$ may be given
approximately by 
\begin{equation}\label{taua}
\tau_\infty \simeq \tau_a = min \; \{ - \frac{l^I_0}{K^I} \}
\; \; .
\end{equation}
Also, $\tau_a \;$ is maximum, and $\tau_{a, \; max} = \frac{1}
{K} \;$, when $K^1 = x^1 \;$, $K^2 = x^2 \;$, $K^3 = min \; \{
x^1 + x^2, \; x^3 \} \;$ and $K^4 = min \; \{ x^1 + x^2, \;
\frac{1}{2} \; (x^1 + x^2 + x^3), \; x^4 \} \;$ where $x^I = -
l^I_0 K \;$, equation (\ref{41a}) at $\tau = 0 \;$ determines $K
> 0 \;$, and we assume with no loss of generality that $ 0 < x^1
\le \cdots \le x^4 \;$. No explicit solution is needed to
evaluate $\tau_a \;$ and $\tau_{a, \; max} \;$.

We studied several sets of $(l^I_0, \; K^I) \;$ numerically and
obtained $\tau_{\infty, \; max} \;$, the maximum of $\tau_\infty
\;$, by sampling $25000 \;$ random sets of $K^I \;$ for each set
of $l^I_0 \;$. We find that $l^I \;$ all diverge at finite $\tau
= \tau_\infty \;$ and that, when $e^{l^I_0} \ll 1 \;$ for all $I
\;$, the approximations given above are quite good : $l^I >
l^I_0 + K^I \tau \;$ discernibly only for $\tau \stackrel {>}
{_\sim} \tau_\infty - 4 \;$; $\; \tau_a \sim (0.5 - 1.1) \;
\tau_\infty \;$ generically; and, for $K^I \;$ which maximise
$\tau_a \;$, we get $\tau_a = \tau_{a, \; max} \sim (0.9 - 1.1)
\; \tau_\infty \; \sim (0.9 - 1.1) \; \tau_{\infty, \; max} \;$.

To convey an idea of what values are possible in equation
(\ref{411}), and also an idea of how good the approximations
given above are, we consider two illustrative sets of $l^I_0
\;$, choose $\; K^I \;$ which maximise $\tau_a \;$, choose \\
$L^i = \sqrt{\frac{1}{6}} \; (- 1, 2, 2, - 1, 0, 0, 0, - 1, - 1,
- 1) \;$ so that $g_s \;$ can be small, and choose $u =
\frac{2}{3} \;$ which corresponds to radiation filled universe
in $(3 + 1) \;$ -- dimensions. The corresponding numerical
results are given in Table I, from which $e^{v^1} \;$ and
$e^{v^c} \;$ can be read off easily using equation (\ref{411}).
Also, $(\tau_{a, \; max}, \; \tau_{\infty, \; max}) = (5.27, \;
5.82) \;$ for the first set, and $= (25.43, \; 25.69) \;$ for
the second set of $l^I_0 \;$ in Table I.


\vspace{2ex}

\begin{tabular}{||c||c||c|c|c||} 
\hline \hline 
& & & & \\ 

$\{ - l^I_0 = - ln \; \rho_{I 0} \} \;$
& $\tau_\infty \;$
& $\frac{M_{11}}{M_4} \;$
& $\frac{M_s}{M_4} \;$
& $g_s \;$ 
\\ 
& & & & \\ 
\hline  \hline 

& & & & \\ 
$5, 5, 12, 12 \;$
& $5.73 \;$
& $ 1.67 * 10^{- 2} \;$
& $ 2.17 * 10^{- 3} \;$
& $ 2.17 * 10^{- 3} \;$
\\
& & & & \\ 
\hline  

& & & & \\ 
$ 20, 30, 40, 50 \;$
& $25.64 \;$
& $ 1.92 * 10^{- 7} \;$
& $ 1.97 * 10^{- 12} \;$
& $ 1.09 * 10^{- 15} \;$
\\
& & & & \\ 
\hline 

\hline 

\end{tabular}


\begin{center}

{\bf Table I :} {\em 
The numerical results for $(\tau_\infty; \; \frac{M_{11}}{M_4},
\; \frac{M_s}{M_4}, \; g_s) \;$ for two illustrative sets of
$l^I_0 \;$. Other parameters are chosen as explained in the
text.}

\end{center}


For a given set of $l^I_0 \;$, our choice of $(K^I, \; L^i) \;$
in Table I results in near--minimum values for $(\frac{M_{11}}
{M_4}, \; \frac{M_s}{M_4}, \; g_s) \;$ within about an order of
magnitude. Our numerical studies confirm this. Also note that,
since $E = 1 \;$, $\; \lambda^i_t (0) = k^i \simeq K^I \simeq
L^i \simeq {\cal O} (1) \;$ naturally whereas ensuring that
$\rho_{I 0} = e^{l^I_0} \ll 1 \;$ for all $I \;$ requires (fine)
tuning. Thus, we conclude that our model naturally leads to
$M_{11} \simeq M_s \simeq M_4 \;$ and $g_s \simeq 1 \;$ within a
few orders of magnitude; and that smaller $M_{11} \;$ and $M_s
\;$, for example $M_s \simeq TeV \simeq 10^{- 16} \; M_4 \; $ as
required in Large Volume compactification scenarios \cite{add},
are also possible but require a corresponding fine tuning of
initial values.

\vspace{4ex}

{\bf 6.}  
We have shown that, in our model, three spatial directions
expand and seven directions stabilise to constant sizes $e^{v^i}
\;$, $\; i = 1, \cdots, 7 \;$. We have also given exact
expressions for $v^i\;$, which depend on initial values and
$\tau_\infty \;$. $\tau_\infty \;$ can be evaluated explicitly
if solutions are known, otherwise numerically. Also, we give
approximate expression for $\tau_\infty \;$ which is easy to
evaluate and works well under certain conditions. Explicit
relations among $(M_{11}, \; M_4, \; M_s, \; g_s) \;$ then
follow from which we see, for example, that obtaining $M_s
\simeq TeV \;$ requires fine tuning.

We conclude by listing a few questions of obvious importance for
further studies. {\bf (i)} How to solve equations (\ref{42a}) --
(\ref{neata}) analytically? {\bf (ii)} Is there any way of
obtaining $M_s \simeq TeV \;$ in the present model without fine
tuning? {\bf (iii)} Why $22'55' \;$ configuration and why not,
for example, $22'2'' \;$ (which will lead \cite{k} to four
spatial directions expanding)? The likely answer is that $22'55'
\;$ configuration is entropically favourable \cite{k2, cm, k},
but dynamical details are not clear. {\bf (iv)} What is the
evolution when topology of spatial directions is more general?
{\bf (v)} We pointed out an interesting similarity with black
holes. Does it have any deeper significance?


\vspace{3ex}

{\bf Acknowledgement: } We thank B. Sathiapalan and
N. V. Suryanarayana for their comments.


\end{document}